\documentclass[12pt,a4paper]{article}
\usepackage{amsmath}
\usepackage{amsfonts}
\usepackage{amssymb,multirow,multicol,booktabs}
\usepackage{graphicx,hyperref,float}
\hypersetup{colorlinks = true, linkcolor = blue, anchorcolor = blue, citecolor = blue, filecolor = blue, urlcolor = blue}
\usepackage[top=1.05in,left=1in,right=1in,bottom=1.05in]{geometry}
\newcommand\BibTeX{{\rmfamily B\kern-.05em \textsc{i\kern-.025em b}\kern-.08em
T\kern-.1667em\lower.7ex\hbox{E}\kern-.125emX}}

\begin{document}
	
%\runtitle{Stability Selection for Lasso, Ridge and Elastic Net}

\small{
\begin{center}
		\textbf{\Large Analysis of Unobserved Heterogeneity via Accelerated Failure Time Models Under Bayesian and Classical Approaches}\\	\vspace{.5cm}

%        \textbf{Running Title: Unobserved Heterogeneity via AFT}
\end{center}
	\vspace{.7cm}
	
	\begin{center}
		{Shaila Sharmin\\ Institute of Statistical Research and Training, University of Dhaka, Dhaka-1000, Bangladesh, Email: ssharmin1@isrt.ac.bd}
\\and\\
		{Md Hasinur Rahaman Khan\footnote{the corresponding author, phone: +8801725106661}\\ Institute of Statistical Research and Training, University of Dhaka, Dhaka-1000, Bangladesh, Email: hasinur@isrt.ac.bd}
%		E-mail: $^1$bhossain@isrt.ac.bd, $^2$hasinur@isrt.ac.bd}\\
		\vspace{0.5cm}
		
		\noindent\rule{16cm}{0.4pt}
	\end{center}
%
%\textbf{\large{Total Word Counts: 4100}}\\
%
\newpage
\textbf{\large{Abstract}}\\
This paper deals with unobserved heterogeneity in the survival dataset through Accelerated Failure Time (AFT) models under both frameworks--Bayesian and classical. The Bayesian approach of dealing with unobserved heterogeneity has recently been discussed in Vallejos and Steel (2017), where mixture models are used to diminish the effect that anomalous observations or some kinds of covariates which are not included in the survival models. The frailty models also deal with this kind of unobserved variability under classical framework and have been used by practitioners as alternative to Bayesian. We discussed both approaches of dealing with unobserved heterogeneity with their pros and cons when a family of rate mixtures of Weibul distributions and a set of random effect distributions were used under Bayesian and classical approaches respectively. We investigated how much the classical estimates differ with the Bayesian estimates, although the paradigm of estimation methods are different. Two real data examples--a bone marrow transplants data and a kidney infection data have been used to illustrate the performances of the methods. In both situations, it is observed that the use of an Inverse-Gaussian mixture distribution outperforms the other possibilities. It is also noticed that the estimates of the frailty models are generally somewhat underestimated by comparing with the estimates of their counterpart.

{\bf Keywords}: AFT model; Bayesian; Classical; Jeffrey's prior; Mixture models; Rate mixtures of Weibull distribution
	}\\
\noindent\rule{16cm}{0.4pt}
%\newpage

\section{Introduction}

In recent years, the increasing availability of event history data in biomedical, economics and social science research has led to the wide spread application of continuous-time survival models. When using such methods, researchers often assume that they have measured and included the relevant causal influences in the model. But most scientists will agree that in any particular empirical analysis, this is hardly ever the case. Usually some important factors could not be measured or were ignored, creating a spurious change over time in estimated transition rates. Hence comes the necessity of incorporating unobserved heterogeneity.
\vspace*{12pt}

\noindent Most commonly, survival data are handled by means of the proportional hazards model, which was first introduced by Cox (1972) and was widely known as the Cox regression model. The central objective of this model is to assess the effects on time to event of only observable covariates by estimating their coefficients. Thus, the conventional Cox model does not always provide an adequate fit to the data and then can generates biases and affects variances of the parameter estimates. One of the reasons is due to the omission of relevant covariates representing information that cannot be observed or have not been observed (univariate case). Another reason can be explained by the violation of the traditional assumption that event times are statistically independent and identically distributed when observed covariates are included. In fact, certain individual are linked by criteria that may share several of the above common unobserved factors (multivariate case).
\vspace*{12pt}

\noindent Originally, Vaupel et al. (1979) proposed a random effects model in order to account for unobserved heterogeneity due to unobserved susceptibility to death. In their studies, the concept of frailty has been introduced and applied in univariate survival models. Their purpose behind introducing the random frailty effect to a Biostatistics framework was to improve the ﬁt of mortality models in a given population.
\vspace*{12pt}

\noindent  Early work on the mixture survival models was done by Berkson and Gage (1952). Two major classes of models have been developed in this context: parametric mixture models based on the standard failure time densities (log-normal, exponential, Weibull, Gompertz etc) using maximum likelihood (ML) to estimate cured proportion and death rate, and non-parametric methods based on the Kaplan-Meier empirical distribution function estimator or on its generalizations. Simulation studies have also been proposed by Ghitany et al.~(1995) to evaluate the properties of ML estimators, in order to assess sample size and follow-up length or to compare the performances of parametric and non-parametric methods in estimating the immune proportion. Hougaard (1995) discussed on frailty models for censored data. He noticed that when gamma distribution for the frailty is used, there is a restriction that the dependence is most important for late events. More generally, the distribution can be stable, inverse Gaussian, or follow a power variance function exponential family. Theoretically, large differences are seen between the choices. Many other researchers worked on unobserved heterogeneity or frailty in survival analysis field (e.g., Aalen, 1987, 1992; Liu, 2014).
\vspace*{12pt}

\noindent Hutton and Solomon (1997) introduced a one-parameter family mixture survival model including both the accelerated life and the proportional hazards models, and they considered the implications for the robustness of estimators of regression coefficients. Gustafson (2001) addressed inference about a single estimator in the context of miss-specification of statistical models in general, using large sample and small coefficient assumptions. As examples, he considered univariate survival analysis models and concluded that the mean and variance are robust, but quantiles are not. Very limited number of works has been done with unobserved heterogeneity under Bayesian framework. Recently, Vallejos and Steel (2017) developed a Bayesian approach for outlying observations and unobserved heterogeneity for AFT models by introducing a family of rate mixtures of Weibull distributions in the model framework. They constructed a weakly informative prior that combined the structure of the Jeffreys prior with a proper prior on the parameters
of the mixing distribution, which was induced through a prior on the coefficient of variation. The improper prior was shown to lead to a proper posterior distribution under some mild conditions. Their methodology mitigates the effect of extreme observations and provides outlier detection method by exploiting the mixing structure.
\vspace*{12pt}

\noindent We discuss both classical and Bayesian approaches of dealing with with unobserved heterogeneity in the data. We mainly discuss classes of survival distributions that provide a natural way to deal with unobserved heterogeneity under Bayesian as proposed in Vallejos and Steel (2017) and existing classical approaches. This analysis allows an arbitrary random effect distribution and focuses on only Accelerated Failure Time (AFT) models, where the interpretation of the regression coefficients is unaffected by the choice of mixing distribution. Bayesian inference,as proposed in Vallejos and Steel (2017), is considered with such models under different weakly informative improper prior distributions. By combining the structure of the Jeffreys prior with a proper (informative) prior, elicited through the coefficient of variation, very mild and easily verified conditions for posterior existence are provided. The appropriateness of different mixing distributions is assessed using standard Bayesian model comparison methods. The covariate effects estimated under classical approach are used for comparison purpose.
\vspace*{12pt}

\noindent The main purpose of this research is to provide an overview of frailty as the effect of unobserved heterogeneity in survival models under Bayesian and classical approaches. Under the Bayesian framework Rate Mixtures of Weibull distributions, for which a random effect at subject level, is introduced through the rate parameter while covariates are incorporated through AFT models. We then investigate the general frailty effects for the same data under classical approach and finally make comparison between the results of classical and Bayesian approaches.

\section{Methodology}
\subsection{Unobserved Heterogeneity}
Ordinary survival models deal with the simplest case of independent and identically distributed data. This is based on the assumption that the study population is homogeneous. But it is a basic observation of medical statistics that individuals differ greatly. So do the effects of a drug, or the influence of various explanatory variables. This heterogeneity is often referred to as variability and it is generally recognized as one of the most important sources of variability in medical and biological applications. This heterogeneity may be difficult to assess, but it is nevertheless of great importance.
\vspace*{12pt}

\noindent The unit considers survival models with a random effect representing unobserved heterogeneity of frailty, a term first introduced by Vaupel et al. (1979). Standard survival models assume homogeneity: all individuals are  subject to the same risks embodied in the hazard $\lambda(t)$ or the survivor functions $S(t)$. Models with covariates relax this assumption by introducing observed sources of heterogeneity. So we can consider unobserved sources of heterogeneity that are not readily captured by covariates, also when individuals have different frailties.
\vspace*{12pt}

\subsection{Frailty Distributions}
\noindent An estimate of the individual hazard rate without taking into account the unobserved frailty may underestimate the hazard function to an increasingly greater extent as time goes by. To be aware of such selection effects, mixture models could be used. That means the population is assumed to be a mixture of individuals with (at least partly unknown) different risks. The non-observable risks are described by the mixture variable, which is called frailty. It is a random variable, which follows some distribution. Different choices of distributions for the unobserved covariates are possible, including binary, gamma and log-normal, which show both qualitative and quantitative differences.
\vspace*{12pt}

\noindent In view of connecting the hazard $\lambda(t)$ and $E(\theta|T \ge t)$, the expected frailty of survivors to $t$ under gamma frailty must be $$E(\theta|T\ge t) = \frac{1}{1+\sigma^2 \Lambda_0(t)}.$$ In fact, we can obtain the whole distribution of frailty among survivors to t. The conditional density of $\theta$ given $T \ge t$ is $ g(\theta| T\ge t) = \theta^{\alpha-1}e^{-(\beta+\Lambda_0(t))^\theta}(\beta + \Lambda_0(t))^\alpha / \Gamma(\alpha)$; a gamma density with parameters $\alpha$ and $\beta+\Lambda_0(t).$
\vspace*{12pt}

\noindent Another distribution that can be used to represent frailty is the inverse Gaussian distribution. Under the multiplicative frailty model the distribution of $\theta$ among survivors to $t$ is also inverse Gaussian, with parameters $\alpha$ and $\beta + \Lambda_0(t)$. In particular, the mean frailty of survivors is $$E(\theta|t \ge t) =   \sqrt{\frac{\alpha}{\beta + \Lambda_0(t) }} =\frac{1}{(1+2\sigma^2\Lambda_0(t))^{1/2}}.$$ Interestingly, the distribution of frailty among those who die at $t$ is a ``generalized" inverse Gaussian.
\vspace*{12pt}

\noindent Log-normal frailty models are especially useful in modelling dependence structures in multivariate frailty models. However, the log-normal distribution has also been applied in univariate cases, for example by Flinn and Heckman (1982) and two variants of the log-normal model exist. We assume a normally distributed random variable $W$ to generate frailty as $Z = e^W$. The two variants of the model are given by the restrictions $EW = 0$ and $EZ = 1$, where the first one is much more popular in the literature. Unfortunately, no explicit form of the unconditional likelihood exists. Consequently, estimation strategies based on numerical integration in the maximum likelihood approach are required.
\vspace*{12pt}

\subsection{Rate Mixtures of Weibull Distributions}
\noindent The Weibull distribution is routinely applied in survival analysis. Its flexibility allows for both increasing and decreasing hazard rates. Since $T_i$ be a positive-valued random variable distributed as a Rate Mixture of Weibull distributions its density function is given as

\begin{equation}\label{eq1:test}
f(t_i|\alpha,\gamma,\theta)=\int\limits_{0}^{\infty} \gamma\alpha\lambda_it_i^{\gamma-1}\exp^{-\alpha\lambda_it_i^{\gamma}}dP_{\Lambda_i}(\lambda_i|\theta), t_i,\alpha,\gamma >0, \theta\in \Theta,
\end{equation}
\vspace*{12pt}
\noindent where $\lambda_i$ is a realization of a random variable $\Lambda_i \sim P_{\Lambda_i}(.|\theta)$. Denote this by $T_i \sim RMW_p(\alpha,\gamma,\theta)$ and it can be represented as

\begin{equation*}
T_i|\alpha,\gamma,\Lambda_i = \lambda_i \sim Weibull(\alpha\lambda_i, \gamma), \hspace{.5cm}
\Lambda_i|\theta~P_{\Lambda_i}(.|\theta).
\end{equation*}

\noindent When $\gamma$ =1, Vallejos and Steel (2017) referred this as the Rate Mixtures of Exponentials (RME) family which is denoted by $T_i\sim RME_p(\alpha, \theta)$. The RME case can be extended to the RMW family via a power transformation if $T_i \sim RME_p (\alpha, \theta)$ then $T_i^{1/\gamma} \sim RMW_p(\alpha,\gamma,\theta)$. If $\gamma\leq 1$, the hazard rate induced by the mixture decreases regardless of the mixing distribution. For $ \gamma > 1$, it has a more flexible shape and can accommodate non-monotonic behavior. The mixing distribution can, in principle, correspond to any proper probability distribution. However, some restrictions are required for identifiability reasons so that we may impose some identifiability conditions for $(\alpha,\gamma,\theta)$, which will be imposed for inference throughout.
\vspace*{12pt}

\noindent Table \ref{tab} displays some examples in the RME family and this list can be extended by selecting other mixing distributions. All these examples generalize to the RMW case via the power transformation mentioned earlier.

\begin{table}[H]
\begin{center}
\caption{Examples in the RME family, $\Theta = (0,\propto)$.}
\vspace{.5cm}
\scalebox{0.8}{\begin{tabular}{ccccc}
\hline
Mixing density&$E(\Lambda_i|\theta)$&$f(t_i|\alpha,\theta)$&$h(t_i|\alpha,\theta)$\\
\hline
Exponential(1)&1&$\alpha(\alpha t_i +1)^{-2}$&$\alpha(\alpha t_i)^{-1}$\\
Gamma$(\theta,\theta)$&1&$\alpha([\alpha/\theta]t_i+1)^{-(\theta+1)}$&$\alpha([\alpha t_i+1])^{-1}$\\
Inv-Gauss$(\theta,1)$&$\theta$&$\alpha e^{(1/\theta)}[{(1/\theta^2)}+2 \alpha t_i]^{-1/2}e{-[(1/\theta^2)+2\alpha t_i]^{1/2}}$&$\alpha[1/\theta^2+2\alpha t_i]^{-1/2}$\\
Log-normal$(0,\theta)$&$e^{\theta/2}$&$(\alpha/\sqrt{2\pi\theta})\int\limits_{0}^{\propto}e^{-\alpha\lambda_i t_i}e^{-(log(\lambda_i))^2/2\theta}d\lambda_i$&No closed form\\
\hline
\end{tabular}}
\label{tab}
\end{center}
\end{table}

\begin{figure}[H]
\centering
\includegraphics[width= 10cm]{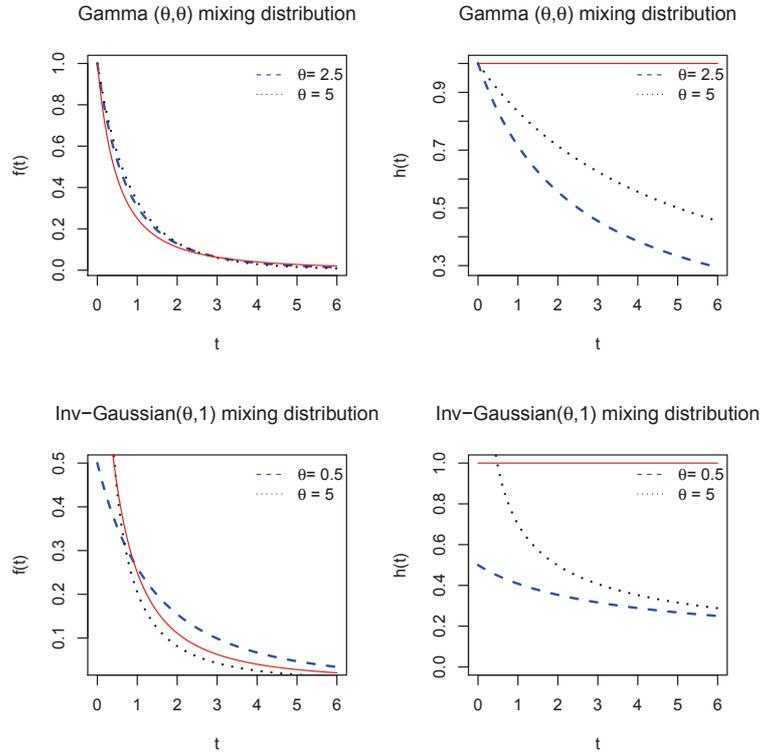}
\caption{Density and hazard function (left and right panels, respectively) of some RME models $(\alpha=1)$. The solid line is the exponential(1) density (hazard).}
\label{fig1}
\end{figure}

\noindent Figure $\ref{fig1}$ as given below shows the RME densities produced by these examples for different values of $\theta$. The density is decreasing (like in the exponential case) but the tail behaviour is very flexible. Figure \ref{fig1} also illustrates that the hazard function decreases over time but that its gradient varies among the different mixing distributions.

\noindent If all the following expressions are well-defined, the coefficient of variation (cv) (i.e. the ratio between the standard deviation and the expected value) of the survival distribution in \eqref{eq1:test}

\begin{equation}\label{eq1:test1}
cv(\gamma,\theta)=\sqrt{\frac{\Gamma(1+2/\gamma)}{\Gamma^2(1+1/\gamma)}\underbrace{\frac{Var_{\Lambda_i}(\Lambda_i^{-1/\gamma}|\theta)}{E_{\Lambda_i}^2(\Lambda_i^{-1/\gamma}|\theta)}}+ \underbrace{\frac{\Gamma(1+2/\gamma)-\Gamma^2(1+1/\gamma)} {\Gamma^2(1+1/\gamma)}}}.
\end{equation}

\noindent The expression in \eqref{eq1:test1} simplifies to $\sqrt{\frac{Var_{\Lambda_i}(\Lambda_i^{-1/\gamma}|\theta)}{E_{\Lambda_i}^2(\Lambda_i^{-1/\gamma}| \theta)}+1}$ when $\gamma=1$. Above equation indicate that $cv(\gamma,\theta)$ is an increasing function of $cv^{*}(\gamma,\theta)$, which is the coefficient of variation of $\Lambda_i{-1/\gamma}$ given $\theta$. In addition, for the same value of $\gamma$, the coefficient of variation of the Weibull distribution $cv^W(\gamma)$ is a lower bound for $cv(\gamma,\theta)$ and they are equal if and only if $\Lambda_i=\lambda_0$ with probability 1. Therefore, evidence of unobserved heterogeneity can be quantified in terms of the ratio

\begin{equation*}
{R_{cv(\gamma,\theta)}}=\dfrac{cv(\gamma,\theta)}{cv^{W}(\gamma)}
\end{equation*}

\noindent defined as the inflation that the mixture induces in the coefficient of variation (w.r.t. a Weibull model with the same $\gamma$). If $\theta$ is such that $cv^*(\gamma,\theta)$ goes to zero, then $R_{cv}(\gamma,\theta)$ tends to one and the mixture reduces to the underlying Weibull model itself. If $\gamma \rightarrow 0, cv^W(\gamma)$ and, consequently, $cv(\gamma,\theta)$ become unbounded. In that case, $R_{cv}(\gamma,\theta)$ behave as $\sqrt{[cv^*(\gamma,\theta)]^2+1}$. If $\gamma=1$, then $R_{cv}(\gamma,\theta)$. Throughout the rest of this paper, we restrict the range of $(\gamma,\theta)$ such that $cv$ is finite (this restriction is not required when $\theta$ does not appear). This facilitates the implementation of Bayesian inference.

\subsection{Regression Model for the RMW Family}
Let $x_i$ be a vector containing the value of $k$ covariates associated with the survival time $i$ and $\beta\in \mathbb{R}^ K$ be a vector of parameters. In the RMW-AFT model, the covariates affect the time scale through the parameter $\alpha$. This model is defined as

\begin{equation*}
T_i \sim RMW_p(\alpha_i,\gamma, \theta), \quad \alpha_i = e^{-\gamma x_i\prime \beta}, i= 1, \dots, n, \text{or equivalently}
\end{equation*}
\hspace{0.5cm}
\begin{equation*}
log(T_i) = x_i\prime \beta + log (\Lambda_i^{1/\lambda}T_0), \quad \text{with} \quad \Lambda_i|\theta \sim P_{\Lambda_i}(\theta) \quad \text{and} \quad T_0|\gamma \sim Weibull(1,\gamma).
\end{equation*}

\noindent The RMW-AFT is itself an AFT model with baseline survival function given by the distribution of $T_0^{\prime}= \Lambda_i^{-1/\gamma}T_0,T_0^{\prime}|\theta\sim RMW_p(1,\gamma,\theta)$. Under this model, $e^{\beta_j}$ can be interpreted as the proportional marginal change of the lifetime distribution percentiles (e.g., median) after a unit change in covariate $j$.

\subsection{Bayesian Inference for the RMW-AFT Model}
The inference procedure is proposed in Vallejos and Steel (2017), where they firstly defined a prior for the RME-AFT model (i.e. fixing $\gamma=1$). In the absence of prior information, a popular choice is to use priors based on the Jeffreys rule, which require the Fisher information matrix (FIM). Jeffreys-style priors can be expressed as $\Pi (\beta, \theta)\propto \Pi(\theta),$  where $\pi(\theta)$ is the part of the prior that depends on $\theta$. As the role of $\theta$ is specific to each mixture, improper priors for $\theta$ will not allow the comparison between models in the RME family using Bayes factors.
\vspace*{12pt}

\noindent Vallejos and Steel (2017) proposed a simplification of the Jeffreys-style priors. They kept the structure, $ \Pi (\beta, \theta)\propto \Pi(\theta)$, where $\pi(\theta)$ is the part of the prior that depends on $\theta$ but assign a proper $\pi(\theta)$. The comparison between the models given below is meaningful if, regardless of the mixing distribution, $\pi(\theta)$ reflects the same prior information (i.e. the priors are ``matched"). Vallejos and Steel (2017) achieved this by exploiting the relationship between $\theta$ and $cv$, the coefficient of variation of the survival times. A proper prior, which is common for all models, is proposed for $cv$ and denoted by $\pi^*(cv)$ that only provides information about $\theta$. The functional relationship between $cv$ and $\theta$ for some distributions in the RME family is derived here.

\begin{table}[H]
\begin{center}
\caption{$cv^*(\gamma,\theta)$ and its derivative w.r.t. $\theta$ for some RMW models. $\Theta = (0,\infty)$, unless otherwise stated and $\psi(.)$ is the digamma function}
\vspace{.5cm}
\begin{tabular}{ccc}
\hline
Mixing & $ [cv^*(\gamma, \theta)]^2 $ & $|\frac{d[cv^*(\gamma,\theta)]^2}{d\theta}|$ \\
\hline
%Exponential(1)&1&$\alpha(\alpha t_i +1)^{-2}$&$\alpha(\alpha t_i)^{-1}$\\
Gamma $(\theta,\theta),\theta>(2/\gamma)$ & $ \frac{\Gamma(\theta)\Gamma(\theta-2/\gamma)}{\Gamma^2(\theta-1/\gamma)} -1$ & $   \frac{\Gamma(\theta)\Gamma(\theta-2/\gamma)}{\Gamma^2(\theta-1/\gamma)}[\psi(\theta)+\psi(\theta-2/\gamma)-2\psi(\theta-1/\gamma)] $ \\
Inverse-Gaussian$(\theta,1)$ & $\frac{\Gamma(\theta)\Gamma(\theta+2/\gamma)}{\Gamma^2(\theta+1/\gamma)}-1 $ & $ \frac{\Gamma(\theta)\Gamma(\theta+2/\gamma)}{\Gamma^2(\theta+1/\gamma)}[\psi(\theta)+\psi(\theta+2/\gamma)-2\psi(\theta+1/\gamma)]$\\
Log-normal$(0,\theta)$ & $ e^{\theta/\gamma^2}-1$ & $ \frac{1}{\gamma^2}e^{\theta/\gamma^2} $\\
\hline
\end{tabular}
\label{tab2}
\end{center}
\end{table}

\noindent Using Equation \eqref{eq1:test1}, the functional relationship between $cv$ and $\theta$ for some distributions in the RME family is also derived. The induced prior for $\theta$ is then easily derived by a change of variable. When comparing a model with $\theta$ to models without $\theta$, meaningful results derive from the fact that the prior on $\theta$ is reasonable. Conditional on $\gamma$, we define $\pi(\theta|\gamma)$ as in the RME-AFT case (via a prior for $cv$, $\pi^*(cv)$). Using $cv(\gamma,\theta)$ and $cv(\gamma,\theta)$ and $cv^*(\gamma,\theta)$ as defined in Equation \eqref{eq1:test1}:

\begin{equation*}
\pi(\theta|\gamma)=\pi^*(cv(\gamma,\theta))\big|\frac{dcv(\gamma,\theta)}{d\theta}\big|,
\textit{where}
\frac{dcv(\gamma,\theta)}{d\theta}=\frac{\Gamma(1+2/\gamma)}{\Gamma^2(1+1/\gamma)}\frac{1}{2cv(\gamma,\theta)}\frac{d[cv^*(\gamma,\theta)]^2}{d\theta}.
\end{equation*}

\noindent Let $T_1,\dots,T_n$ be the survival times of $n$ independent individuals. Let observe survival times are defined by $t_1,\dots,t_n$ and define $X = (x_1 \dots x_n)^{\prime}$. Assume that $n\geq k, X$ has rank $k$ (full rank) and that the prior for $(\beta,\gamma,\theta)$ is proportional to $\pi(\gamma,\theta)$ which is a proper density function for $(\gamma,\theta)$. If $t_i \neq 0$ for all $i = 1, \dots ,n$, the posterior distribution of $(\beta,\gamma,\theta)$ is proper. A proper prior for $(\gamma,\theta)$ is used so that it assures a proper posterior distribution if $X$ has full column rank and there are no zero observations of the survival time. Posterior propriety can be precluded when conditioning on a particular sample of point observations. However, point observations do not affect the posterior propriety for the RMW-AFT model.
\vspace*{12pt}

\noindent Mixing parameters are handled through data augmentation and we implement an adaptive Metropolis-within-Gibbs sampler with Gaussian random walk proposals. As the Weibull survival function has a known simple form, the data augmentation is not used for dealing with censored observations. The full conditionals for the Gibbs sampler are given by

\begin{eqnarray}
	\nonumber
	\pi(\beta_j|\beta_{-j},\gamma,\theta,\lambda, t,c) & \propto & e^{-\gamma \beta_j \sum_{i=1}^{n} c_i x_{ij}} e^{-\sum_{i=1}^{n} \lambda_i (t_i e^{- x_i^\prime \beta} )^\gamma }\\ \nonumber
	\pi(\gamma|\beta,\theta,\lambda,t,c) & \propto &  \gamma^{\sum_{i=1}^{n}c_i}  \left[\prod_{i=1}^{n} t_i^{c_i}\right]^{\gamma-1}  e^{-\gamma \sum_{i=1}^{n} c_i x_i^\prime \beta} e^{-\sum_{i=1}^{n} \lambda_i (t_i e^{- x_i^\prime \beta} )^\gamma} \pi(\theta| \gamma)\pi(\gamma)\\ \nonumber
	\pi(\theta|\beta,\gamma,\lambda,t,c) & \propto & \prod_{i=1}^{n} dP(\lambda_i|\theta) \pi(\theta|\gamma)\\	 \nonumber
	\pi(\lambda_i|\beta,\gamma,\theta,\lambda^{-i},t,c) & \propto & \lambda_i^{c_i} e^{-\lambda_i(t_i e^{-x_i\prime \beta})^\gamma} dP(\lambda_i|\theta),  i= 1,\ldots ,n
	\end{eqnarray}
where $\beta_{-j}=(\beta_1,\dots ,\beta_{j-1},\beta_{j+1},\beta_{k})$, $\lambda_{-i}=\lambda_1,\dots,\lambda_{i-1},\lambda_{i+1},\lambda_n$ and the $c_i$'s, $i = 1,\dots ,n$ are censoring indicators equal to $1$ if the survival time for individual $i$ is observed and $0$ if it is censored. For a general mixing distribution, Metropolis updates are required in all full conditionals. Nevertheless, Gibbs steps can be used for particular mixing distributions.
\vspace*{12pt}

\noindent The adequacy of a particular mixing distribution is evaluated using standard Bayesian model comparison criteria: Bayes factors (BF), deviance information criteria (DIC) and pseudo Bayes factors (PsBF). The BF between two models is the ratio between the marginal likelihoods and for each observation $i$, the BF are computed for the model $M_0 : \Lambda_i = \lambda_ref$ versus $M_1 : \Lambda_i \neq\lambda_ref$(all other $\lambda_j,j\neq i$free), where $\lambda_ref$is a reference value (specific to the mixing distribution). The BF in favour of $M_0$ versus $M_1$ can be computed as

\begin{equation*}
BF_{01}^{(i)}=\pi(\lambda_i|t,c)E\bigg(\frac{1}{dp(\lambda_i|\theta)}\bigg)\bigg|_{\lambda_i=\lambda_ref}=E\bigg(\frac{\pi(t_i|\beta,\gamma,\theta,\lambda_i,c_i)dp(\lambda_i|\theta)}{\pi(t_i|\beta,\gamma,\theta,c_i)}\bigg)E\big(\frac{1}{dp(\lambda_i|\theta)} \bigg)\bigg|_{\lambda_i=\lambda_ref},
\end{equation*}

\noindent where the expectation are w.r.t. $\pi(\beta,\gamma,\theta|t,c)$ and $\pi(\theta|\Lambda_i=\lambda_ref,t,c)$, respectively.
\vspace*{12pt}

\noindent The DIC is defined as $DIC\equiv E(D(\beta,\gamma,\theta,t)|t)+P_D$, where $D(\beta,\gamma,\theta,t)=-2log(f(t|\beta,\gamma,\theta))$ (deviance function) and $P_D=E(D(\beta,\gamma,\theta,t)|t)-D(\hat{\beta},\hat{\gamma},\hat{\theta},t)$ (effective number of parameters) with $\hat{\beta},\hat{\gamma}$ and $\hat{\theta}$ being the posterior medians of $\beta,\gamma$ and $\theta$, respectively. DIC is computed using the marginal model (after integrating the $\lambda_i$'s), and lower values suggest better models.

\section{Analysis and Results}
\subsection{Autologous and Allogenic Bone Marrow Transplant}
\noindent This dataset (Klein and Moeschberger, 1997) contains post-surgery information about 101 advanced acute myelogenous leukemia patients. The endpoint of the study is the disease-free survival time, i.e. until relapse or death (in months). The disease-free survival time was observed for 50 patients while the others are right-censored. In the trial, 51 patients received an autologous bone marrow transplant. This replaces the patient's marrow with their own marrow after the application of high doses of chemotherapy. Only the type of treatment is available as a covariate and thus an important amount of unobserved heterogeneity is expected.
\vspace*{12pt}

\noindent The standard graphical check of $log(-log(s(t)))$ versus $t$ is checked here which suggests that the PH assumption does not hold. This graph should result in parallel curves if the predictor is proportional. Even the residual exhibits a random (i.e. unsystematic) pattern at each failure time, then this gives evidence the covariate effect is not changing with respect to time precisely the PH assumption. As we got this residual plot where the errors are systematic, it suggests that as time passes, the covariate effect is changing. Now we can conduct our analysis with parametric model.

\begin{figure}[H]
\centering
\includegraphics[width=.4\linewidth]{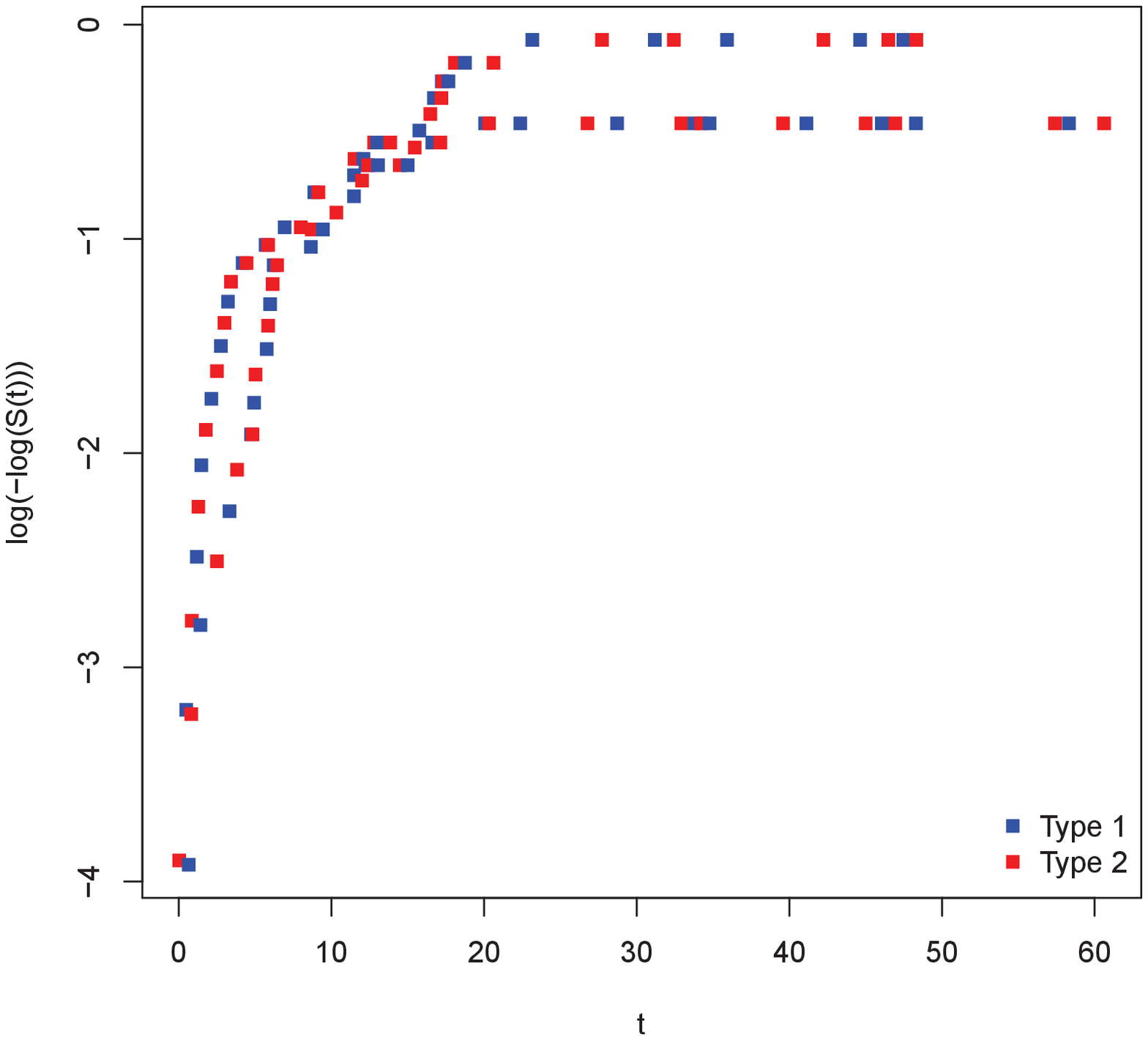}
\includegraphics[width=.4\linewidth]{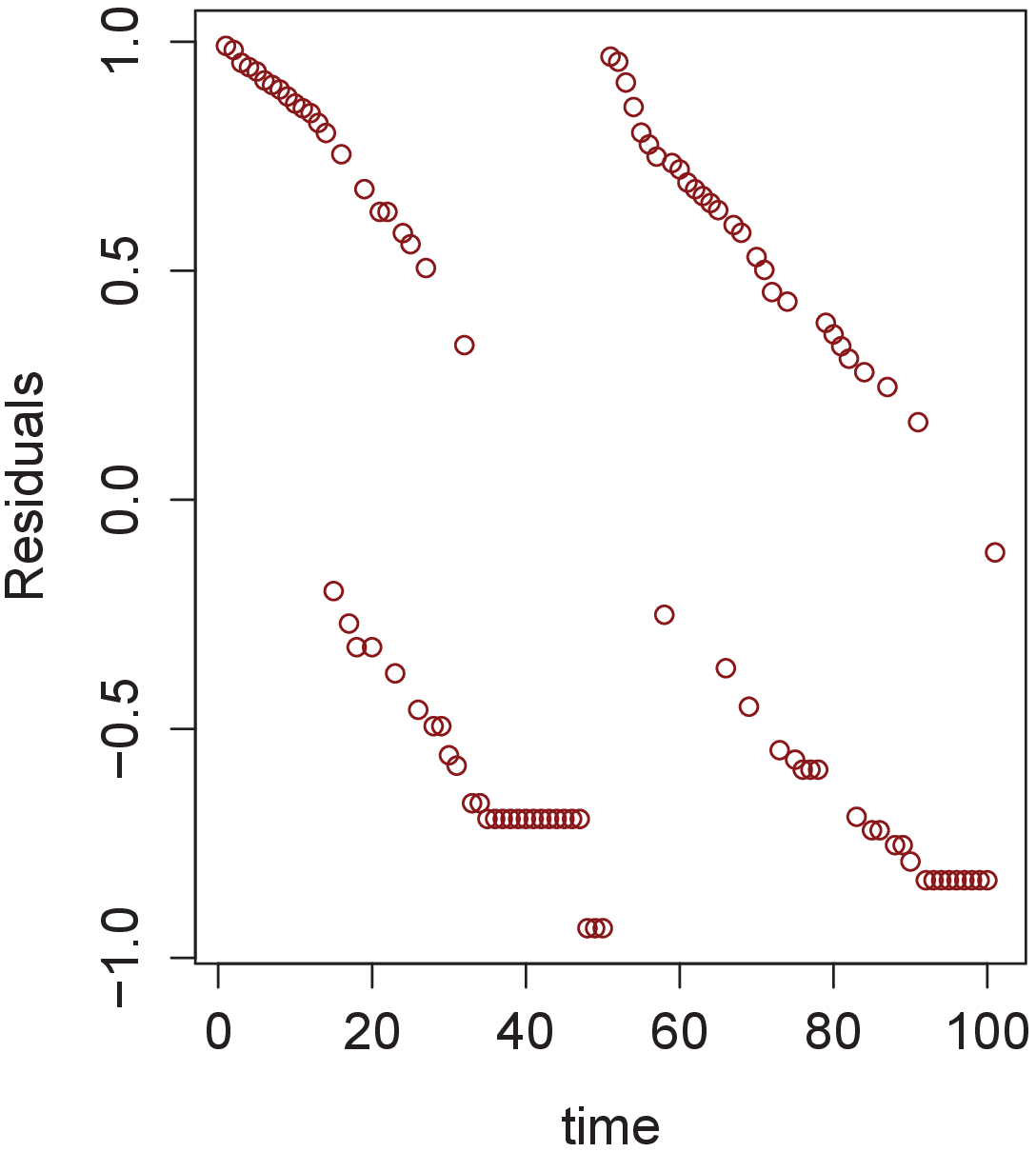}
\caption{Checking Proportional Hazards Assumption and Residual Plot for Cox Model}
%\label{fig:inter}
\end{figure}

\noindent The data is first analysed using exponential and Weibull AFT models. In addition, RME and RMW-AFT models with the mixing distribution in Table 1 are fitted using the priors proposed here. For all models, the total number of iteration is 600,000. After burn-in of 25\% of the initial iterations and thinning, the following results are presented on the basis of 9,000 draws.In contrast to the Weibull case and Exponential(1) presented in Table \ref{tab} there is evidence of mixing for all the RMW- AFT regressions. We obtain far apart points for the mixing densities rather than the Exponential and Weibull which are the models without mixing. Based on this evidence, RME-AFT models are used for this data. We adopt $E(cv)$ equals to 1.5, 2, 5 and 10(if there is no $\theta$ in the model, all these priors coincide). Large values of $E(cv)$ are associated with stronger prior beliefs about the existence of unobserved heterogeneity.

\begin{figure}[H]
\centering
\includegraphics[width=.4\linewidth]{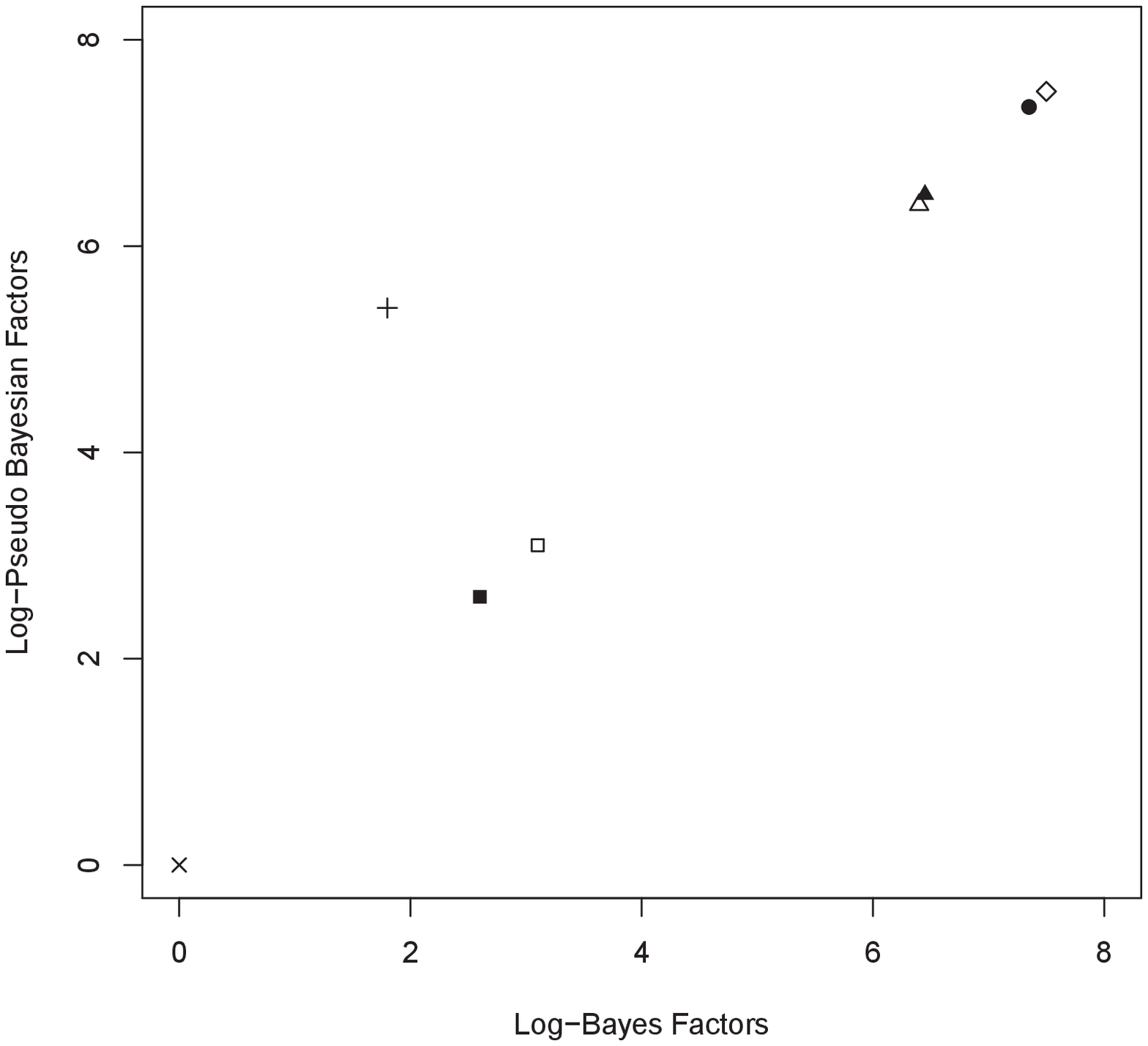}
\includegraphics[width=.4\linewidth]{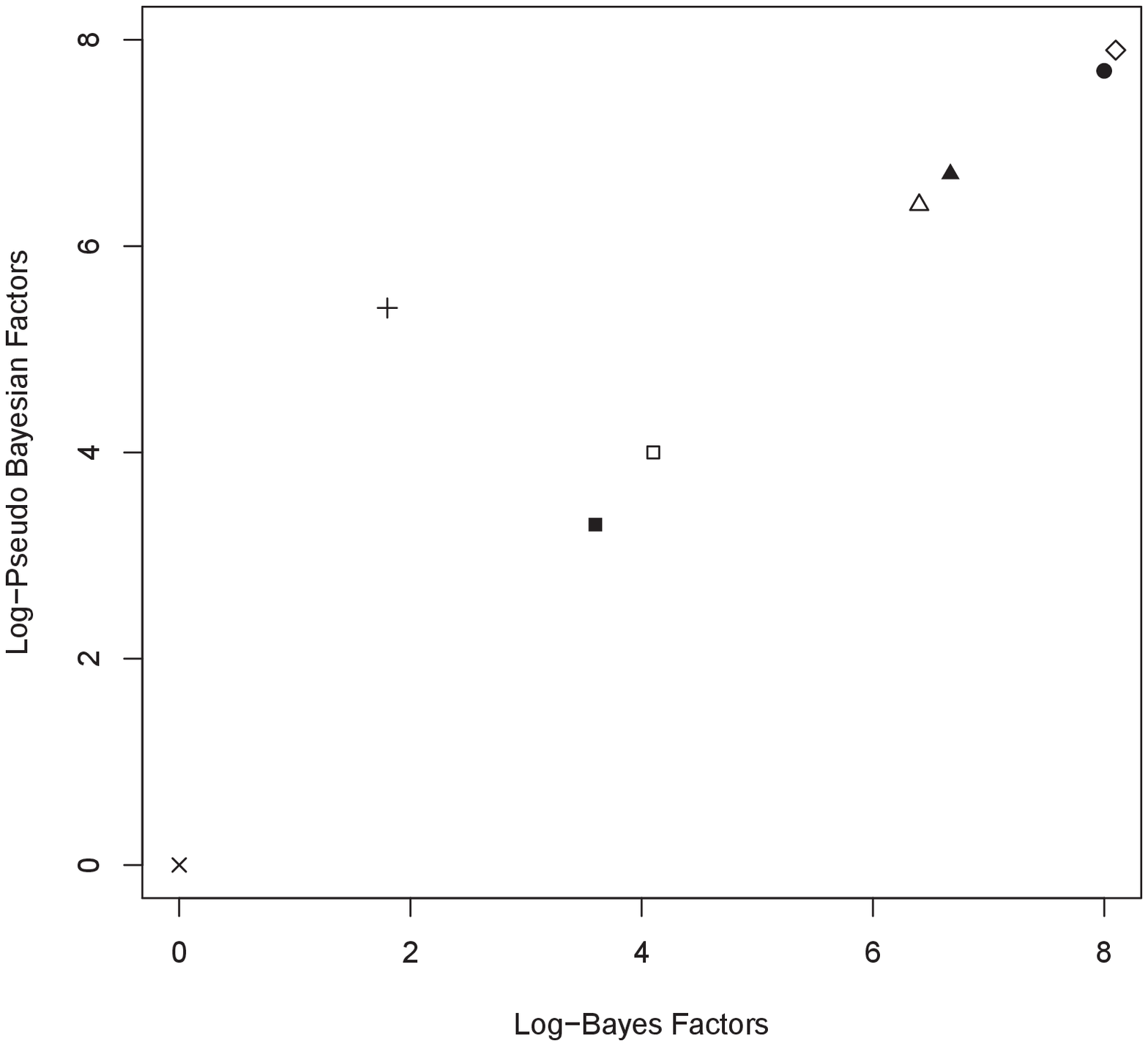}
\includegraphics[width=.4\linewidth]{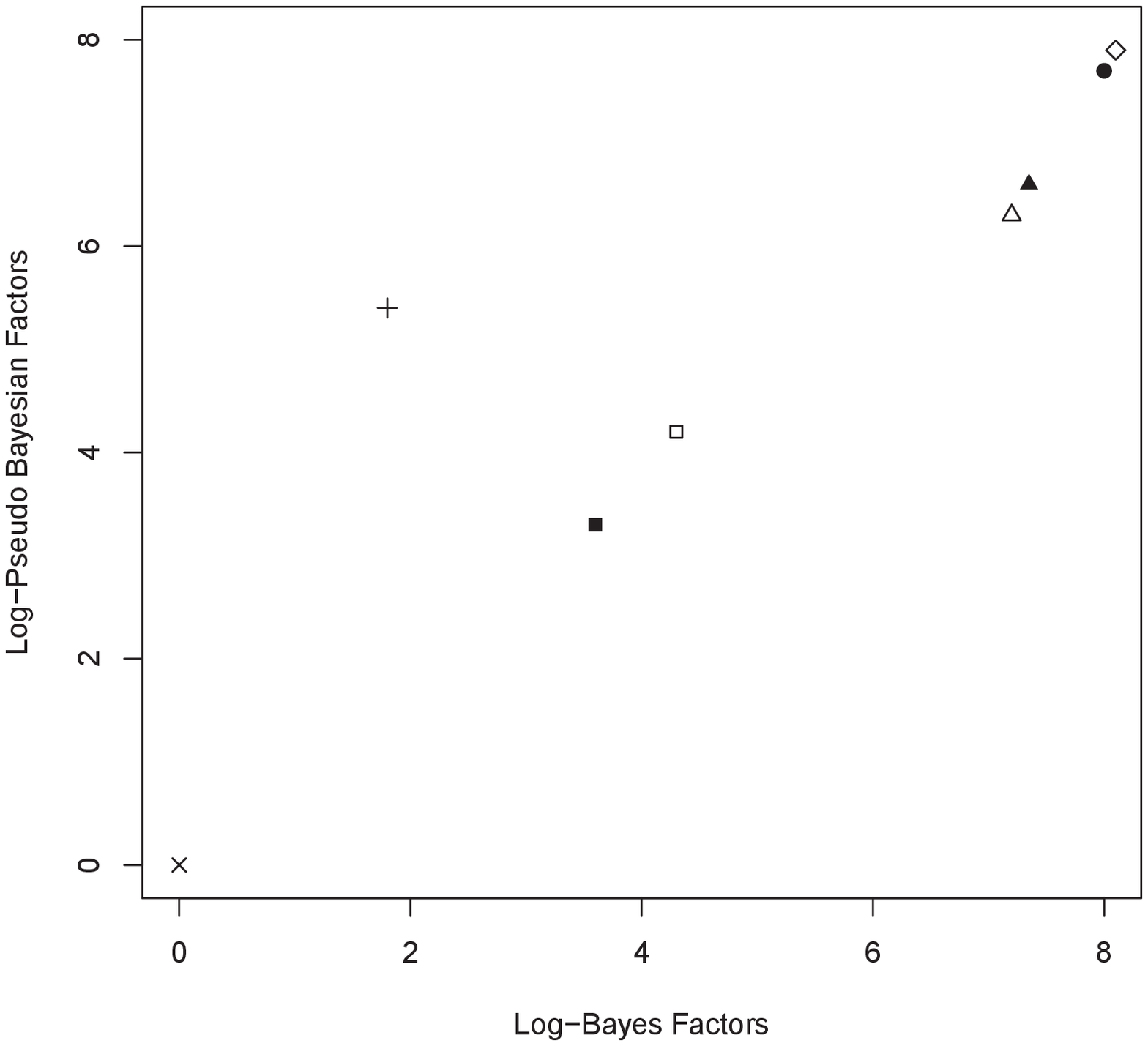}
\includegraphics[width=.4\linewidth]{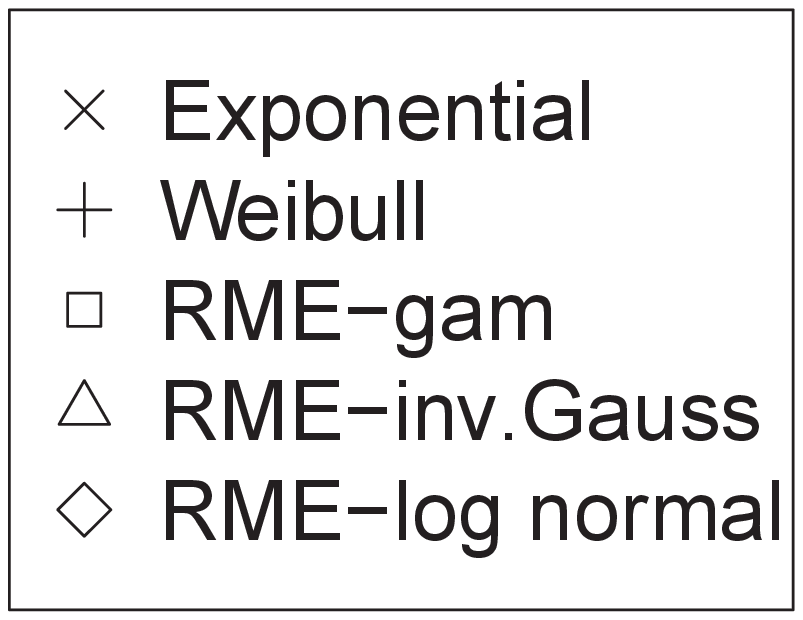}
\caption{Identifying mixing and Model Comparison in terms of Bayes Factor and Pseudo Bayes Factor with $E(cv)=2,5,10$ from left to right}
%\label{fig:inter}
\end{figure}

%\pagebreak
\noindent The presence of unobserved heterogeneity is supported by the data. Figure 3 compares the fitted models in terms of BF and PsBF (w.r.t. the exponential model). For all priors considered, all the mixture models in Table \ref{tab} support over the exponential model. The Weibull model is also beaten in terms of BF and PsBF. The Pseudo-BF is a predictive criterion and is virtually unaffected by these changes in prior. The similarity of both criteria for the mixture models is indicative of the fact that priors are well-matched.
\vspace*{12pt}

\noindent DIC also suggests a similar ordering between models. DIC is computed using the marginal model and lower
values suggest better models. The Inverse-Gaussian mixing receives most support overall. It is easy to implement that the log-normal mixing distribution has slightly more support for large E(cv), but rather less for small E(cv). Interestingly, the popular gamma mixing is the least preferred of all mixing distributions. This suggests the need for a mixture and is consistent with strong heterogeneity in the data that leads to support for the inverse gaussian mixing model.
\vspace*{12pt}

\noindent Now we may represent the effects in both purpose of Bayesian and classical approach. We have the coefficient effects along with the frailty value i.e. the term for random effect. In Bayesian approach, we have reported 95\% highest posterior density interval and in Classical approach, we have reported 95\% confidence interval.

\begin{table}[H]
\caption{Bayesian and Classical Frailty Effects for Bone Marrow Data}
\centering
\scalebox{0.80}{
\begin{tabular}{ccccccc}
\hline
Model & \multicolumn{3}{c}{Bayesian Approach} & \multicolumn{3}{c}{Classical Approach}  \\
 & \multicolumn{3}{c}{Estimate(95$\%$ HPD Interval)} & \multicolumn{3}{c}{Estimate (95 $\%$ CI)} \\ \hline

 & $\beta_0$ & $\beta_1$ & Frailty & $\beta_0$ & $\beta_1$ &  Frailty \\ \hline
Exponential & 3.079 & -0.061 & & 2.95  &-0.078 &   \\
  &  (2.51,3.66)& (-0.821,0.687) & & (2.16,3.75)&(-0.003,0.18) & \\ \hline
Weibull & 3.76 & -0.325 & &3.76 &-0.425 & \\
  & (-0.881,0.247)& (-0.091, 1.35) & &(3.08,4.13)&(-0.13,1.2)& \\  \hline
  RME-GAM & 3.521 & -0.263 &2.86& 3.52&-0.274&1.26\\
 &(-3.042,4.04)&(-0.911,0.368)&(2.11,6.24)&(3.13,3.96)&(-0.08,0.19)&(1.07,2.11)\\ \hline
  RME-IG & 4.231 & -0.228 & 4.08&3.27&-0.239&2.44\\
  &(2.73,5.37)& (-1.127,0.611) &(1.43,7.23) &(2.19,4.01)&(-0.15,0.24)&(2.13,3.27)\\ \hline
  RME-LN & 3.64 & -0.135 & 3.04 & 3.48 & -0.178 & 2.12\\
  &(3.125,4,16)&(-0.841,0.611)&(0.414,2.76)&(3.15,4.53)&(-0.113,0.15)&(1.86,3.01)\\ \hline
\end{tabular}
}
\label{table_bayes_clas}
\end{table}

\noindent In classical context, the models with random effect are suggested to use since the frailty estimates lie within their 95\% confidence interval. Furthermore, the frailty estimates and in some cases the parameter estimates are shown to be underestimated by comparing with their Bayesian estimates. Moreover, in Bayesian purpose the highest posterior density can summarize the uncertainty by giving a range of values on the posterior probability distribution with 95\%. This suggests the need for a mixture and is shown to be consistent with strong heterogeneity in the data that leads to support for the inverse gaussian mixing model. Whereas the choice of a prior affects inference on $R_{cv}$, the posterior distribution of $\beta$(usually the parameter of interest) is more robust.

\subsection{Kidney Catheter Data}

This dataset (McGilchrist and Aisbett, 1991) on the recurrence times to infection, at the point of insertion of the catheter, for kidney patients using portable dialysis equipment. The data consist of 38 patients with two recurrence times given for each with the covariates age and gender. Catheters may be removed for reasons other than infection, in which case the observation is censored. Each patient has exactly 2 observations. Hence the individual specific frailty is expected. Here we have the follow-up time along with the censoring status. The recurrence time measures the time between catheter insertion and infection, which occurs where the catheter is inserted. When infection occurs, the catheter is removed and the infection is treated, and then, after a pre-determined period of time, the catheter is reinserted. When the catheter is removed for reasons other than infection, the time to infection is treated as censored.
\vspace*{12pt}

\begin{figure}[H]
\centering
\includegraphics[width=.7\linewidth]{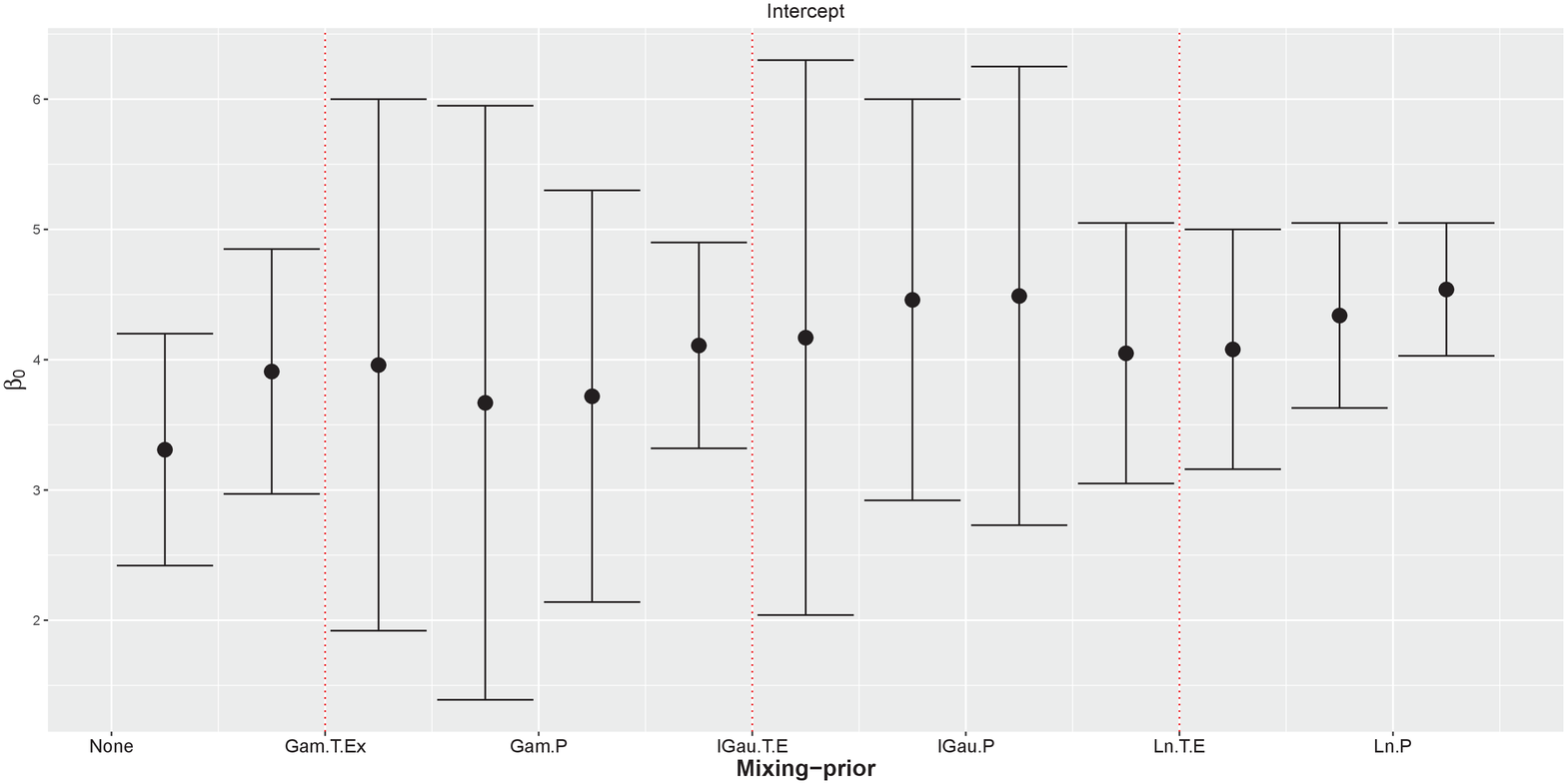}
\includegraphics[width=.7\linewidth]{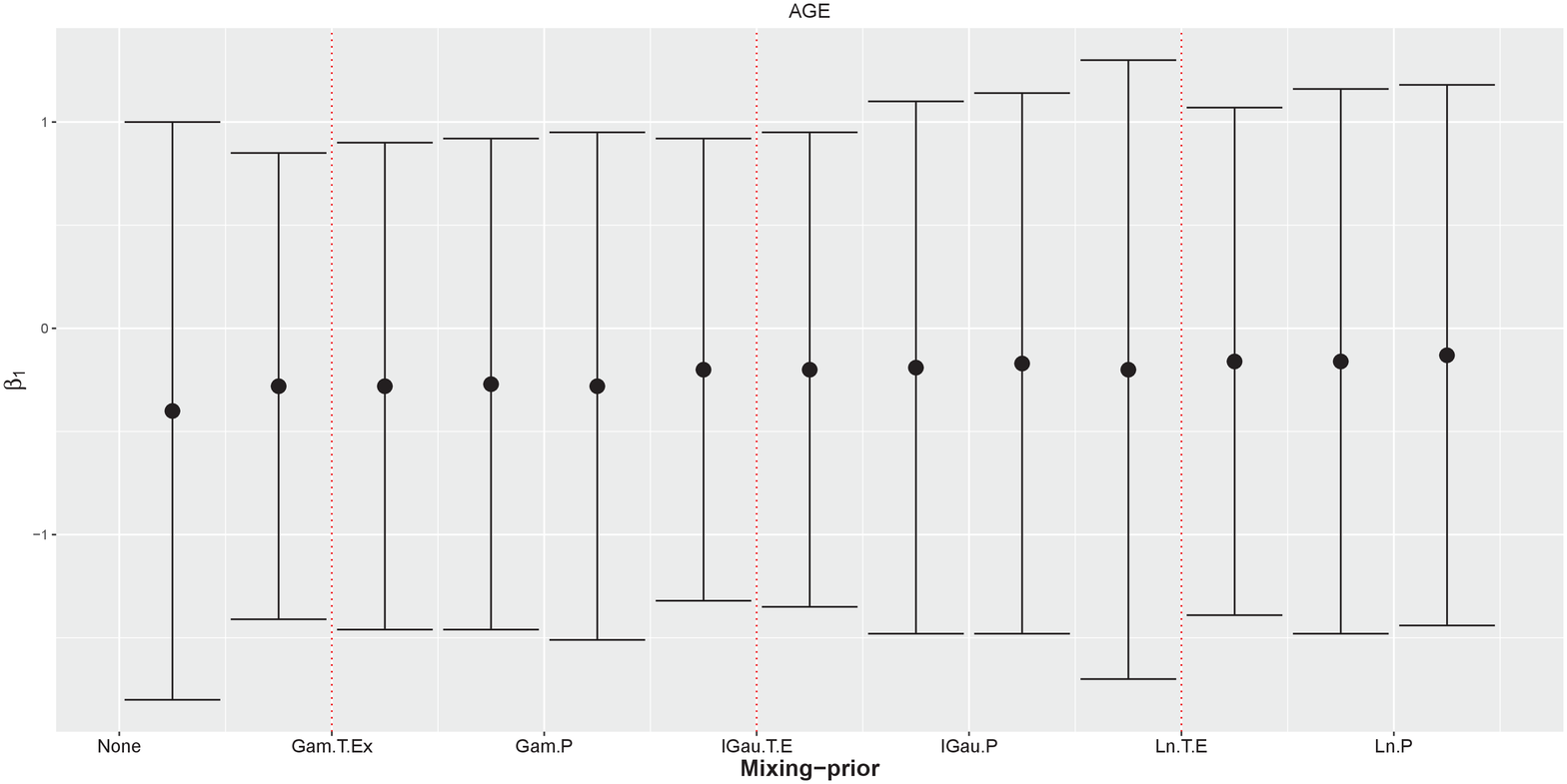}
\includegraphics[width=.7\linewidth]{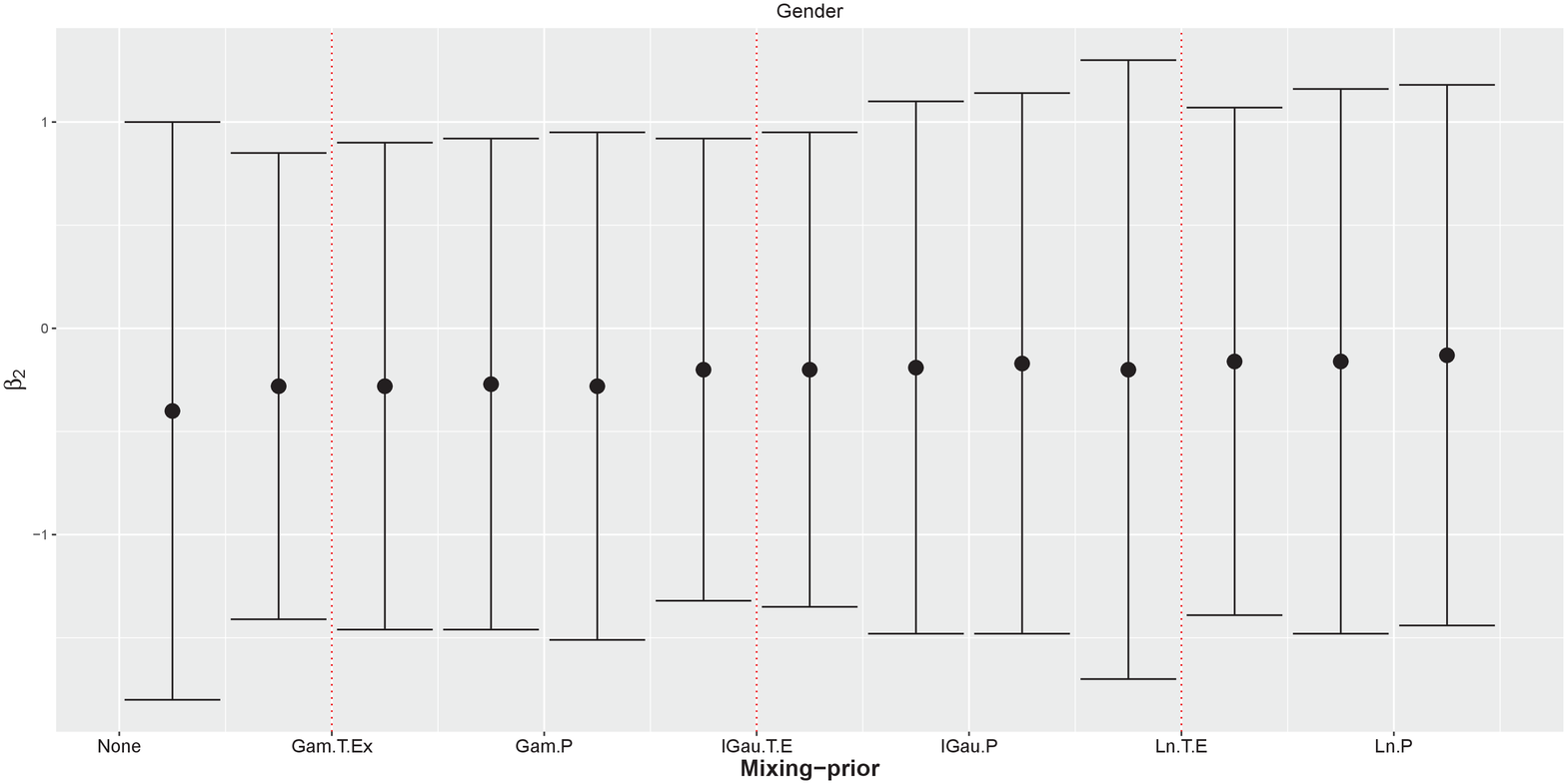}
\caption{Posterior of the regression coefficients, $\beta_0, \beta_1, \beta_2$ (from top to bottom) for Kidney Catheter data using the RME-AFT model}
\label{fig:inter}
\end{figure}

\noindent The data are analyzed using the RMW-AFT model with the mixing distributions in Table \ref{tab}. For comparison, a Weibull regression is also fitted. Figure \ref{fig:inter} shows that mixture models estimate a similar effect of the covariates. The effect of covariates ($\beta_0, \beta_1, \beta_2$) is with the Weibull model that is without mixing and also with other mixing model. Here the priors can be justified and all the median values lie within the 95\% HPD interval.
\vspace*{12pt}

\noindent Model comparison in terms of Bayes Factors and Pseudo Bayes Factors (with respect to the Weibull model) for the mixing distributions analysed here. the mixture models provide a better fit for the data and lead to better predictions. In fact, for all priors considered, all the mixture models have a better performance in terms of BF and PsBF. Again, both criteria are very close.
\vspace*{12pt}

\noindent In addition, the DIC criteria arranged models in the same order. The result strongly suggests the existence of unobserved heterogeneity which is also strong for the inverse-gaussian mixing. This evidence supports that the posterior distribution of $R_cv$ is concentrated away from one. Overall, the Inverse-Gaussian mixing provides the best results in terms of BF, PsBF and DIC.

\begin{table}[H]
\begin{center}
\caption{Model Comparison in terms of Deviance Information Criteria(DIC) for Kidney Data }
\vspace{.5cm}
\begin{tabular}{cc}
\hline
Model & DIC Value\\
\hline
Weibull & 395.281\\
Rate Mixture Exponential with Gamma Mixing & 351.536\\
Rate Mixture Exponential with Inverse-Gaussian Mixing & 345.179\\
Rate Mixture Exponential with Log-normal Mixing & 349.028 \\
\hline
\end{tabular}
\end{center}
\end{table}

\noindent As lower values of DIC suggest better model, Inverse-Gaussian Mixing of Rate Mixture Exponential receives most support over all other models. Now the tabulated values show the frailty effects for both Bayesian and classical purposes.

\begin{table}[H]
\centering
\caption{Bayesian and Classical Frailty Effects for Kidney Data}
%\resizebox{\textwidth}{!}{
	\scalebox{.7}{

\begin{tabular}{ccccccccc}
\hline
Model & \multicolumn{4}{c}{Bayesian Approach} & \multicolumn{4}{c}{Classical Approach}  \\
&  \multicolumn{4}{c}{Estimate(95$\%$ HPD Interval)} & \multicolumn{4}{c}{Estimate (95 $\%$ CI)} \\ \hline

 & $\beta_0$ & $\beta_1$ & $\beta_2$ & Frailty & $\beta_0$ & $\beta_1$ & $\beta_2$ & Frailty \\ \hline
Weibull &3.31  & -0.009 & 1.42  &   &3.61 &-0.15  &1.52  & \\
& (2.93,4.17)& (-0.002,1.15)  &(1.13,2.59) & &(2.29,4.84) &(-0.038,0.016) &(0.344,2.33) & \\ \hline
RME-GAM &3.71  & -0.17 & 1.45  & 1.56  &3.65 &-0.009  &1.26  & 1.44\\
& (2.71,4.53)& (-0.002,1.139)  &(1.05,2.78) & (0.09,2.01)&(3.01,4.17) &(-0.06,1.04) &(1.12,2.37) &(0.359,5.733) \\
\hline
RME-IG & 3.92& -0.15& 1.37 &3.59& 3.65& -0.009& 1.26& 1.13 \\
&(3.02,5.01)& (-0.05,0.29)& (1.11,2.74)& (1.53,4.67)& (2.22,5.08)& (-0.038,0.019)& (0.169,2.35)& (0.737,1.74)\\ \hline
RME-LN &3.81& -0.19& 1.35& 2.06& 3.53& -0.151& 1.39& 1.09 \\
&(3.049,4.92) &(-0.05,1..53)& (1.01,3.04) &(1.023,3.25)& (2.08,3.89) &(-0.058,0.027)& (0.59,2.01) &(0.08,1.73) \\ \hline
\end{tabular}
}
\label{table_bayes_clas}
\end{table}

\noindent All those values in classical approach lie within 95\% confidence interval. Also the Highest Posterior Density  gives a range of values on the posterior probability distribution with 95\% for the coefficients and the frailty term which shows the effectiveness of the random effect. It is also evident from this data analysis that the frailty estimates along with the parameter estimates, in most cases, are shown to be much underestimated by comparing with their Bayesian estimates.

\section{Discussion and Conclusion}
Mixtures of life distribution are proposed in order to account for unobserved heterogeneity in survival datasets. In particular, the family generated by mixtures of Weibull distributions with random rate parameter is explored in detail (and its special case of rate mixtures of exponentials). These mixtures are shown to induce a larger coefficient of variation than the original Weibull distribution and more flexible hazard functions. Vallejos and Steel (2017) implemented a family of such mixture models and found also that they induced a larger coefficient of variation than the original Weibull distribution.
\vspace*{12pt}

\noindent In our analysis, all of those models with random effect are justified under both classical and Bayesian frameworks. For both approaches, the HPD intervals and also the confidence intervals contain the frailty terms. Both practical dataset analysis provides strong evidence for unobserved heterogeneity existing in the datasets. We also have noticed as Vallejos and Steel (2017) that mixture models are supported by the data properly in terms of Bayes factors, Pseudo-Bayes Factor and Deviance Information Criteria value. In particular, the use of an Inverse-Gaussian mixture distribution leads to the overall best results among both of the applications. It is evident from the analysis that even the frailty effect exists in the classical models just as the Bayesian approach.
\vspace*{12pt}

\noindent It is also evident from the analysis that inference on the regression coefficients is relatively robust to the prior assumptions and even to the choice of mixing distribution. The main differences in the estimates are observed between the mixture models and the Weibull models under both of the data sets. So we can say that the mixture models diminish the effect that anomalous observations have over posterior inference. It is worth of mentioning that significant difference is observed between the Bayesian and classical esitmates for both analysis. Particularly, the estimates under classical approach have been significantly underestimated as compared with their classical approaches. We don't know the reasons why this happens but we may suggest that the classical approach may fail to quantify the total uncertainty of the unobserved heterogeneity which, we believe can be properly possible to achieve under Bayesian framework by suitably choosing the priors to account for the extra uncertainty due to unobserved heterogeneity in the data sets.

%\section*{Conflict of Interest}	
%The authors have no conflict of interest.

%\newpage
%\phantom{aaaa}
\end{document}